\documentclass[journal]{IEEEtran}
\IEEEoverridecommandlockouts
\usepackage[dvipsnames]{xcolor}  

\usepackage{hyperref}
\hypersetup{
    colorlinks=false,           
    linkcolor=black,            
    citecolor=black,            
    urlcolor=black,             
    pdfborder={0 0 2},          
    linkbordercolor=BrickRed,   
    citebordercolor=ForestGreen,
    urlbordercolor=white        
}
\usepackage{microtype}
\usepackage{amsmath,amssymb,amsfonts}
\usepackage{algorithmic}
\usepackage{algorithm}
\floatname{algorithm}{Alg.} 
\usepackage{graphicx}
\usepackage{cleveref}
\crefname{figure}{Fig.}{Figs.}
\crefname{subfigure}{Fig.}{Figs.}
\crefname{algorithm}{Alg.}{Algs.}
\crefformat{figure}{Fig.~#2#1#3}
\crefformat{subfigure}{Fig.~#2#1#3}
\usepackage{svg}
\usepackage{caption}
\usepackage[labelformat=parens,labelsep=space]{subcaption}
\usepackage{url}
\usepackage{stfloats}
\usepackage{textcomp}
\usepackage{xcolor}
\usepackage{url}
\usepackage{verbatim}
\usepackage{array, booktabs}
\usepackage{bm}

\usepackage{xcolor}
\usepackage[
  backend=biber,    
  style=ieee,       
  sorting=none      
]{biblatex}
\appto{\bibfont}{\small}
\def\BibTeX{{\rm B\kern-.05em{\sc i\kern-.025em b}\kern-.08em
		T\kern-.1667em\lower.7ex\hbox{E}\kern-.125emX}}
	
\begin{document}

\title {Multiple STAR-RISs-Empowered Multi-User Communications with Diversified QoS Provisioning}

\author{
	Junfeng Wang,
         Xiao Tang,
	Jinxin Liu,
	Zhi Zhai,
        Qinghe Du,
	and Naijin Liu
	\thanks{J. Wang, J. Liu and Z. Zhai are with the School of Mechanical Engineering, Xi’an Jiaotong University, Xi’an 710049, China (e-mail:  \protect\url{wjf1242600648@stu.xjtu.edu.cn},   \protect\url{jinxin.liu@xjtu.edu.cn}, \protect\url{zhaizhi@xjtu.edu.cn}).}
	\thanks{X. Tang and Q. Du are with  the School of Information and Communication Engineering, Xi'an Jiaotong University, Xi'an 710049, China (e-mail: \protect\url{tangxiao@xjtu.edu.cn}, \protect\url{duqinghe@mail.xjtu.edu.cn}).}
	\thanks{N. Liu is with the Qian Xuesen Laboratory of Space Technology, China Academy of Space Technology, Beijing 100094, China (e-mail: \protect\url{liunaijin@xjtu.edu.cn}).}
}

	\maketitle
	
	\begin{abstract}
		
        This paper proposes a quality-of-service (QoS)-aware multi-user communication framework facilitated by multiple simultaneously transmitting and reflecting reconfigurable intelligent surfaces (STAR-RISs). The user groups are established based on their QoS requirements specified by the minimum data rate, which is provisioned by the optimized transmission and reflection configurations of the STAR-RISs. Particularly, we formulate an optimization problem to maximize the aggregate link rate across all users, under group-specified rate requirements by jointly considering the transmit beamforming and STAR-RIS configurations. Then, we employ the Lagrangian duality with quadratic transformation to tackle the non-convexity of the objective. We decompose the problem within a block coordinate descent framework, and the subproblems are solved through convex approximation and iterated to approach the optimal solution. Simulation results demonstrate the effectiveness of the proposed method in enhancing the system sum rate with guaranteed QoS performance for heterogeneous users, offering valuable insights for the deployment of STAR-RISs in future QoS-aware wireless networks.

	\end{abstract}
	
	
	\section{Introduction}
 
	The evolution of the sixth-generation (6G) wireless communication system heralds a new era of ubiquitous connectivity and highly diverse user experiences. Towards this vision, the intelligent reconfiguration of the wireless propagation environment is crucial  \cite{ref0}. Reconfigurable intelligent surfaces (RISs), as an emerging technology, with its advantages of low power consumption, low cost, and flexible deployment, improve signal quality, enhance coverage, and increase spectral efficiency through passive beamforming, and have become one of the key enabling technologies for 6G \cite{ref2}. RIS intelligently constructs communication channels by regulating the phase of incident electromagnetic waves, effectively enhancing the target signal and suppressing interference.
	
	Despite its promising prospects, traditional RIS is limited by the ``only reflection'' mechanism and often operates solely in the half-space of the incident wave. This inherent limitation restricts its coverage capability, especially in complex environments where users may be located on both sides of the surface or in areas blocked by direct or reflected paths\cite{ref3-1}. To break this limitation and fully exploit the potential of intelligent surfaces in comprehensive spatial coverage, simultaneously transmitting and reflecting reconfigurable intelligent reflecting surfaces (STAR-RISs) have recently been proposed \cite{ref4}. A STAR-RIS can achieve full-space control of electromagnetic waves and has the ability to simultaneously transmit and reflect signals, significantly expanding the system design freedom and providing new possibilities for supporting multi-region and multi-scenario communications \cite{ref5}.
	
	Meanwhile, with the evolution of 6G networks and the increasing diversity of service scenarios, future wireless communication systems need to support highly differentiated and personalized data rate quality of service  guarantees \cite{ref6}. Due to the insufficiency of adaptability and resource allocation mechanisms, traditional network architectures often struggle to provide such fine-grained and diverse QoS guarantees. To this end, STAR-RIS technology, with its inherent full-space beamforming capability and reconfigurable electromagnetic response in both transmission and reflection modes, offers an attractive solution \cite{ref10}. Particularly, by intelligently guiding energy, STAR-RIS can flexibly allocate different data rates to users in the reflection and transmission spaces to meet diversified QoS requirements \cite{ref8-1}. Moreover, the coordinated deployment of multiple STAR-RISs enhances system flexibility in delivering scenario-aware, user-oriented services, further strengthening its ability to support heterogeneous QoS demands. These features position STAR-RIS as a key enabler of customized service provisioning in future wireless systems, and highlight the necessity of developing efficient optimization strategies to fully exploit its potential.

	However, most existing works focus on transmission enhancement, while the potential of leveraging multiple STAR-RISs to realize user-centric, QoS-aware service provisioning remains largely unexplored \cite{ref11}. Towards this issue, this paper investigates a multi-user communication system assisted by multiple STAR-RISs and proposes a framework that jointly optimizes the base station's transmit beamforming and the passive beamforming of multiple STAR-RISs, aiming to maximize the aggregate link rate to enhance overall performance while satisfying the minimum rate requirements of individual users. To address the non-convexity of the formulated problem, we developed an effective iterative solution framework. This approach successfully transforms the original problem into a series of more tractable subproblems, each of which can be reliably solved to achieve a high-quality overall solution. Simulation results validate the effectiveness of the proposed method in enhancing aggregate link rate while ensuring diversified QoS support, demonstrating superior performance compared to baseline schemes.
	
	\section{System Model And Problem Formulation}
	\subsection{System Model}

\begin{figure}[t]
		\centerline{\includegraphics[width=0.4\textwidth]{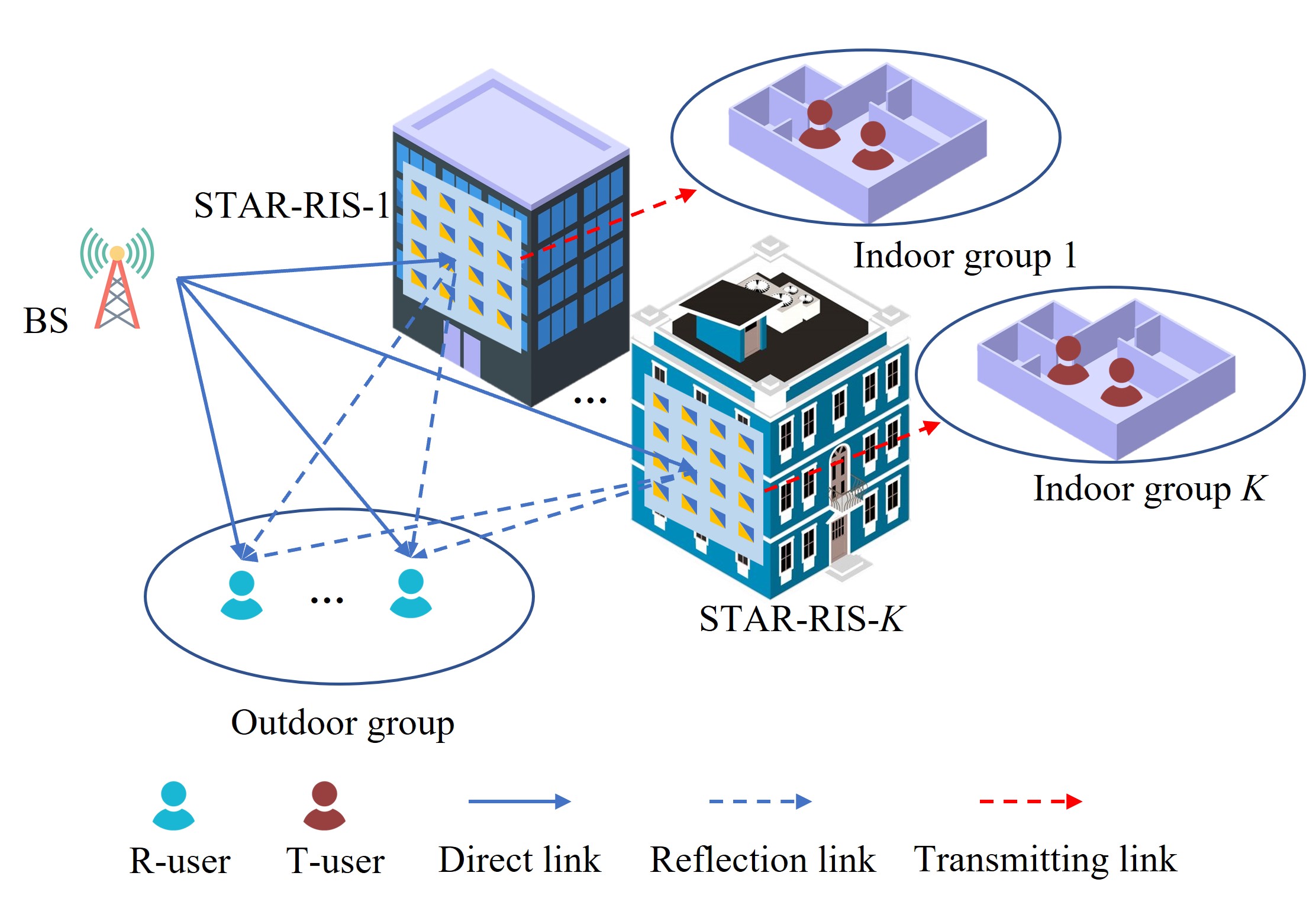}}
		\caption{Multi-STAR-RIS assisted multi-user communication with QoS awareness (illustrated by indoor/outdoor scenario).}
		\label{fig:system}
	\end{figure}

    We consider a wireless network that a base station (BS), equipped with $N_0$ antennas, serves multiple users, which are divided into several groups based on their QoS requirements. To enable effective QoS provisioning, we introduce multiple STAR-RISs to enhance the performance. Particularly, there are $K$ STAR-RISs deployed, denoted by \(\mathcal{K} = \{1, \dots, K\}\), where the \(k\)-th RIS consists of $N_k$ elements. Then, the transmission side of each STAR-RIS covers a group of users, which conduct multi-cast communications with identical QoS requirement specified by the minimum data rate. We denote the user set through the transmission of $k$-th STAR-RIS as \(\mathcal{J}_k = \{1, \dots, J_k\}\), with a required date rate as $R_k^{\min}$. Meanwhile, the rest users in the system, denoted by \(\mathcal{J}_0 = \{1, \dots, J_0\}\), are served in an uni-cast manner by the BS and collective reflection of all the STAR-RISs, each user $j_0$ is associated with the minimum data rate of $R_{j_0}^{\min}$. Note a typical use case of the considered system is the mixed indoor/outdoor scenario, where the the outdoor users correspond to the set $\mathcal{J}_0$, and there are $K$ indoor groups of users for \(\{\mathcal{J}_k\}_{k\in \mathcal{K}}\), and the users are with diversified QoS provisioning, as shown in \Cref{fig:system}.

	We assume that all channels are line-of-sight (LoS) channels, and both the BS  and all STAR-RISs are equipped with uniform linear arrays (ULAs) aligned along the \(x\)-axis, and the array responses are determined by the relative spatial positions of the transceivers. The channel models are defined as follows. The BS-to-\(k\)-th STAR-RIS channel is denoted by \(\boldsymbol{H}_{0,k} \in \mathbb{C}^{N_k \times N_0}\); the BS-to-\(j_0\)-th reflection user channel by \(\boldsymbol{h}_{0,j_0} \in \mathbb{C}^{N_0 \times 1}\); the \(k\)-th STAR-RIS-to-\(j_k\)-th transmission user channel by \(\boldsymbol{h}_{k,j_k} \in \mathbb{C}^{N_k \times 1}\); and the \(k\)-th STAR-RIS-to-\(j_0\)-th reflection user channel by \(\boldsymbol{h}_{k,j_0} \in \mathbb{C}^{N_k \times 1}\). Here, the subscript \(0\) refers to the BS, \(k \in \mathcal{K}\) indexes STAR-RISs, \(j_k \in \mathcal{J}_k\) indexes transmission users, and \(j_0 \in \mathcal{J}_0\) indexes reflection users. Here, we assume that the BS direct transmissions are available to the reflection-side users in \(\mathcal{J}_0 \), while unreachable to the transmission-side users in \(\{\mathcal{J}_k\}_{k\in \mathcal{K}}\). This is because the transmission-side users suffer from severe penetration loss and structural obstructions that block the direct BS link, so they can only rely on the signals transmitted through the RIS. In contrast, the reflection-side users are in environments without significant obstructions, allowing them to receive both the direct signals from the BS and the signals reflected by the RIS.

	Under the energy splitting (ES) protocol, each STAR-RIS unit can independently control its reflection and transmission responses. For the $n_k$-th element of the $k$-th STAR-RIS, the reflection and transmission energy coefficients are denoted by $\beta_{k,n_k}^{\mathsf r}$ and $\beta_{k,n_k}^{\mathsf t}$, and the corresponding phase shifts by $\theta_{k,n_k}^{\mathsf r}$ and $\theta_{k,n_k}^{\mathsf t}$, respectively. Similar to many existing works \cite{ref-resp-1,ref-resp-2,ref-resp-3}, we adopt the ideal continuous-phase shift model for STAR-RIS elements. Notably, prior studies~\cite{ref-resp-4},~\cite{ref-resp-5} show that low-resolution discrete implementations can closely approach the performance of continuous-phase designs, justifying the use of this assumption for theoretical analysis and algorithm development. Each element satisfies: $\beta_{k,n_k}^{\mathsf s} \ge 0$, $\beta_{k,n_k}^{\mathsf r} + \beta_{k,n_k}^{\mathsf t} = 1$, and $\theta_{k,n_k}^{\mathsf s} \in [0, 2\pi)$, where $\mathsf s \in \{\mathsf r, \mathsf t\}$. The complex amplitude coefficient vector is defined as $\boldsymbol{v}_k^{\mathsf s} = [\sqrt{\beta_{k,1}^s}e^{j\theta_{k,1}^{\mathsf s}}, \dots, \sqrt{\beta_{k,N_k}^{\mathsf s}}e^{j\theta_{k,N_k}^{\mathsf s}}]^{\mathsf  T}$, and the corresponding phase shift matrix is $\boldsymbol{\Theta}_k^{\mathsf s} = \mathrm{diag}(\boldsymbol{v}_k^{\mathsf s})$. The sets of reflection and transmission matrices for all STAR-RIS are denoted as $\boldsymbol{\Theta}^{\mathsf r} = \{\boldsymbol{\Theta}_k^{\mathsf r}\}_{k \in \mathcal{K}}$ and $\boldsymbol{\Theta}^{\mathsf t} = \{\boldsymbol{\Theta}_k^{\mathsf t}\}_{k \in \mathcal{K}}$ respectively.

    In this system model, the indoor transmission side users are geographically concentrated and have homogeneous service demands, while the outdoor reflection side users are distributed and require personalized demands. Therefore, the BS adopts multicast to send data to $K$ indoor user groups $\{\mathcal{J}_k\}_{k\in\mathcal{K}}$ and unicast to serve $J_0$ outdoor users in $\mathcal{J}_0$. Overall, the BS generates a total of $K+J_0$ data streams, with the first $K$ being multicast streams and the last $J_0$ being unicast streams. The BS transmits a total of $K + J_0$ independent data streams $\{d_i\}_{i \in \mathcal{I}}$, where $d_i \sim \mathcal{CN}(0,1)$ and $\mathcal{I} = \{1, \dots, K + J_0\}$. Each stream $d_i$ is precoded by the beamforming vector $\boldsymbol{w}_i \in \mathbb{C}^{N_0 \times 1}$, and the resulting transmit signal is $\boldsymbol{x} = \sum_{i\in\mathcal{I}} \boldsymbol{w}_i d_i$. The total transmission power is constrained by $\sum_{i\in \mathcal{I}} \|\boldsymbol{w}_i\|^2 \le P_T$, where $P_T$ denotes the maximum transmit power at the BS.
	
	Accordingly, the received signals at a transmission user \(U_{k,j_k}\) and a reflection user \(U_{j_0}\) are given by
	
	\begin{subequations}
		\begin{align}
			y_{k,j_k} &= \boldsymbol{h}_{k,j_k}^{\mathsf H} \boldsymbol{\Theta}_k^{\mathsf t} \boldsymbol{H}_{0,k}\boldsymbol{x} + n_{k,j_k}, \\
			y_{j_0} &= \left(\boldsymbol{h}_{0,j_0}^{\mathsf H} + \sum_{k\in\mathcal{K}} \boldsymbol{h}_{k,j_0}^{\mathsf H} \boldsymbol{\Theta}_k^{\mathsf r} \boldsymbol{H}_{0,k}\right)\boldsymbol{x} + n_{j_0},
		\end{align}
	\end{subequations}
	where \(n_{k,j_k}\) and \(n_{j_0}\) represent independent additive white Gaussian noise (AWGN) terms with variance \(\sigma_0^2\).For notational simplicity, we define the effective channels as \(\boldsymbol{g}_{k,j_k}^{\mathsf H} = \boldsymbol{h}_{k,j_k}^{\mathsf H} \boldsymbol{\Theta}_k^{\mathsf t} \boldsymbol{H}_{0,k}\) and \(\boldsymbol{g}_{j_0}^{\mathsf H} = \boldsymbol{h}_{0,j_0}^{\mathsf H} + \sum_{k\in\mathcal{K}} \boldsymbol{h}_{k,j_0}^{\mathsf H} \boldsymbol{\Theta}_k^{\mathsf r} \boldsymbol{H}_{0,k}\). Accordingly, the achievable rates are given by 
    \begin{subequations}
		\begin{align}
        R_{k,j_k} &= \log\left(1+\frac{|\boldsymbol{g}_{k,j_k}^{\mathsf  H}\boldsymbol{w}_k|^2}{\sum_{i\neq k}|\boldsymbol{g}_{k,j_k}^{\mathsf  H}\boldsymbol{w}_i|^2 + \sigma_0^2}\right),\\
        R_{j_0} &= \log\left(1 + \frac{|\boldsymbol{g}_{j_0}^{\mathsf  H}\boldsymbol{w}_{j_0}|^2}{\sum_{i\neq j_0 + K}|\boldsymbol{g}_{j_0}^{\mathsf  H}\boldsymbol{w}_i|^2 + \sigma_0^2}\right).
       	\end{align}
    \end{subequations}

	\subsection{Problem Formulation}
	
	We aim to jointly optimize the BS transmit beamforming vectors $\boldsymbol{w} = \{ \boldsymbol{w}_i \}_{i \in \mathcal{I}}$ and the STAR-RIS reflective and transmissive phase shift matrices $\boldsymbol{\Theta}^{\mathsf r}$ and $\boldsymbol{\Theta}^{\mathsf t}$ to maximize the total system throughput while guaranteeing diversified QoS across user groups. Specifically, each user group is assigned a dedicated rate threshold corresponding to its QoS requirement, subject to the BS transmit power constraint and STAR-RIS limitations, including constraints on amplitude, energy splitting, and phase shift adjustment.
	
	Therefore, we aim to maximize the aggregate link rate defined as the sum of rates across all individual user in the network, acting as a comprehensive objective function. This goal enhances the overall robustness and quality for the diversified user groups, and thus the optimization problem is formulated as
	\begin{subequations}\label{eq:main2}
		\begin{align}
			\max_{\boldsymbol{w},\boldsymbol{\Theta}^r,\boldsymbol{\Theta}^t} \quad 
			& \sum_{j_0 \in \mathcal{J}_0} R_{j_0} + \sum_{k \in \mathcal{K}} \sum_{j_k \in \mathcal{J}_k} R_{k,j_k} \label{eq:main2a} \\
			\mathrm{s.t.} \quad
			& R_{k,j_k} \geq R_k^{\min}, \quad \forall j_k \in \mathcal{J}_k, \forall k \in \mathcal{K}, \label{eq:main2b} \\
			& R_{j_0} \geq R_{j_0}^{\min}, \quad \forall j_0 \in \mathcal{J}_0, \label{eq:main2c} \\
			& \sum_{i \in \mathcal{I}} \|\boldsymbol{w}_i\|^2 \leq P_T, \label{eq:main2d} \\
			& \beta_{k,n_k}^{\mathsf{r}},\beta_{k,n_k}^{\mathsf{t}} \in [0,1],\quad\forall n_k \in \mathcal N_k,  \forall k \in \mathcal{K}, \label{eq:main2e} \\
			& \beta_{k,n_k}^{\mathsf t} + \beta_{k,n_k}^{\mathsf r} = 1,\quad \forall n_k \in \mathcal N_k, \forall k \in \mathcal{K}, \label{eq:main2f} \\
			& \theta_{k,n_k}^{\mathsf r}, \theta_{k,n_k}^{\mathsf t} \in [0,2\pi),\quad\forall n_k \in \mathcal N_k,  \forall k \in \mathcal{K} . \label{eq:main2g}
		\end{align}
	\end{subequations}
    For the formulated problem, the constraints \eqref{eq:main2b}–\eqref{eq:main2c} enforce QoS provisioning for transmission groups and reflection groups users, respectively; \eqref{eq:main2d} ensures BS power budget; \eqref{eq:main2e}–\eqref{eq:main2g} capture STAR-RIS hardware limitations.
    The optimization problem in \eqref{eq:main2} is challenging due to two main factors: (i) the objective function is non-concave, primarily due to the presence of logarithmic functions and fractional expressions in the rate formulation; and (ii) the strong coupling among optimization variables, along with the presence of quadratic terms, further contributes to the non-convexity of the problem.
	\section{QoS-Aware Transmission and Reflection Optimization}
	
	 To address the non-convexity of the optimization problem, we first apply the LDT-QT technique to equivalently reformulate the rate expressions, thereby eliminating the coupling between the logarithmic and fractional structures. Then, within the block coordinate descent (BCD) framework, we incorporate the semidefinite programming (SDP) method to alternately optimize the decomposed subproblems.
\subsection{ Problem Reformulation and Decomposition}
    For notational consistency between the transmission and reflection domains, we define the user set as $\mathcal{U} = \bigcup_{k \in \mathcal{K}} \mathcal{J}_k \cup \mathcal{J}_0$. To eliminate the logarithmic term in the rate expression, the Lagrangian dual transformation (LDT) is applied \cite{ref15}. Specifically, a set of auxiliary variables $\boldsymbol{\alpha} = \{\alpha_u\}_{u \in \mathcal{U}}$ is introduced to reformulate the objective function as

	\begin{equation}
		\begin{aligned}
			&f(\boldsymbol w,\boldsymbol \Theta^{\mathsf s},\boldsymbol \alpha)\\
			=&\sum_{u \in \mathcal U}\left( \log(1+\alpha_{u})+f_{u}(\boldsymbol  w,\boldsymbol \Theta^{\mathsf s}, \alpha_{u})-\alpha_{u}\right), \label{eq:3}
		\end{aligned}
	\end{equation}
	where $f_u(\boldsymbol{w}, \boldsymbol{\Theta}^{\mathsf{s}}, \alpha_u) = \frac{(1+\alpha_u) P_u}{Q_u}$, with $P_u = \left|\boldsymbol{g}_u^\mathsf{H} \boldsymbol{w}_u\right|^2$ and $Q_u = \sum_{i \in \mathcal{I}} \left|\boldsymbol{g}_u^\mathsf{H} \boldsymbol{w}_i\right|^2 + \sigma_0^2$. The equivalence between \eqref{eq:main2a} and \eqref{eq:3} holds if and only if the auxiliary variable $\alpha_u$ satisfies $\alpha_u = \frac{P_u}{Q_u - P_u}$, for all $u \in \mathcal{U}$.
	
    To further eliminate the fractional structure in \eqref{eq:3}, the quadratic transformation (QT) technique is applied \cite{ref14}. This method introduces an auxiliary variable \(\boldsymbol{\eta} = \{\eta_u\}_{u \in \mathcal{U}}\) to equivalently reformulate the original expression as
	\begin{equation}
	\begin{aligned}
	f_u(\boldsymbol{w}, \boldsymbol \Theta^{\mathsf s}, \alpha_u, \eta_u) = 2\eta_u \sqrt{(1+\alpha_u) P_u} - \eta_u^2 Q_u, \label{eq:4}
		\end{aligned}
\end{equation}
	where \(\eta_u\) serves as an auxiliary variable. The equivalence holds if and only if \(\eta_u = \sqrt{(1+\alpha_u)P_u}/Q_u\), \(\forall u \in \mathcal{U}\). Based on the preceding transformations, the objective function can now be expressed as
	\begin{equation}
		\begin{aligned}
			&f(\boldsymbol w,\boldsymbol \Theta^{\mathsf s},\boldsymbol \alpha,\boldsymbol \eta)\\
			=&\sum_{u \in \mathcal U}\left( \log(1+\alpha_{u})+f_{u}(\boldsymbol  w,\boldsymbol \Theta^{\mathsf s},\alpha_{u},\eta_{u})-\alpha_{u} \right). \label{eq:5}
		\end{aligned}
	\end{equation}
    
	Through the above transformations, the logarithmic and fractional structures in the objective function \eqref{eq:main2a} have been effectively eliminated. At this stage, the transmit beamforming, STAR-RIS passive beamforming, and the auxiliary variables introduced by the LDT-QT method are mutually decoupled, allowing problem \eqref{eq:main2} to be alternately solved via decomposition within the subsequent BCD framework.
    
\vspace{-8pt} 

\subsection{Transmission Beamforming }

We first consider optimizing the transmit beamforming while fixing the STAR-RIS passive beamforming and auxiliary variables. To facilitate the optimization, we apply the SDP technique by introducing the variable $\boldsymbol{W} = \{\boldsymbol{W}_i\}_{i\in\mathcal{I}}$, where each $\boldsymbol{W}_i = \boldsymbol{w}_i \boldsymbol{w}_i^\mathsf{H}$, subject to $\boldsymbol{W}_i \succeq \boldsymbol{0}$ and $\text{rank}(\boldsymbol{W}_i) = 1$. The desired signal power is defined as
\begin{equation} \label{eq:Pu}
P_u(\boldsymbol{W}) = 
\begin{cases}
	\operatorname{Tr}(\boldsymbol{W}_k \boldsymbol{G}_{k,j_k}), & \text{if } u = j_k, \\
	\operatorname{Tr}(\boldsymbol{W}_{j_0} \boldsymbol{G}_{j_0}), & \text{if } u = j_0,
\end{cases}
\end{equation}
and the total received power is given by
\begin{equation} \label{eq:Qu}
Q_u(\boldsymbol{W}) = \sum_{i \in \mathcal{I}} \operatorname{Tr}(\boldsymbol{W}_i \boldsymbol{G}_u) + \sigma_u^2,
\end{equation}
where \(\boldsymbol{G}_{k,j_k} = \boldsymbol{g}_{k,j_k} \boldsymbol{g}_{k,j_k}^\mathsf{H}\) and \(\boldsymbol{G}_{j_0} = \boldsymbol{g}_{j_0} \boldsymbol{g}_{j_0}^\mathsf{H}\). Thus, the objective function becomes
\begin{equation}
	\overline{f}(\boldsymbol{W}) = \sum_{u\in\mathcal{U}} \left(2\eta_u\sqrt{(1+\alpha_u)P_u(\boldsymbol{W})} - \eta_u^2 Q_u(\boldsymbol{W})\right). \label{eq:9}
\end{equation}
Similarly, the QoS constraints are equivalently expressed in the SDP form as follows
\begin{equation}
	(2^{R_u^{\min}} - 1) Q_u(\boldsymbol{W}) - 2^{R_u^{\min}}P_u(\boldsymbol{W}) \leq 0, \quad \forall u\in\mathcal{U}, \label{eq:10}
\end{equation}
The power constraint is rewritten in SDP form as
\begin{equation}
	\sum_{i\in\mathcal{I}} \operatorname{Tr}(\boldsymbol{W}_i) \leq P_T. \label{eq:11}
\end{equation}
Therefore, the transmission beamforming subproblem is reformulated as
\begin{subequations}\label{eq:12}
	\begin{align}
		\max_{\boldsymbol{W}} \quad & \overline{f}(\boldsymbol{W}) \label{eq:12a}\\
		\text{s.t.} \quad 
		& \boldsymbol{W}_i \succeq 0,~{\rm rank}(\boldsymbol{W}_i) = 1, \quad \forall i\in\mathcal{I}, \label{eq:12b}\\
		& \eqref{eq:10}, \eqref{eq:11}. \nonumber
	\end{align}
\end{subequations}
The problem is rendered convex by relaxing the rank-one constraint in \eqref{eq:12b}, which is the only source of its non-convexity. To this end, we adopt the semidefinite relaxation technique. The relaxed problem becomes a convex semidefinite programming problem, which can be efficiently solved using standard convex solvers such as CVX. Subsequently, the transmit beamforming vector $\boldsymbol{w}_i$ is recovered by extracting the principal eigenvector of $\boldsymbol{W}_i$. If $\boldsymbol{W}_i$ does not satisfy the rank-one condition, a Gaussian randomization procedure is employed to construct a feasible approximate solution.

\renewcommand{\algorithmicrequire}{\textbf{Input:}}
 \renewcommand{\algorithmicensure}{\textbf{Output:}}
 \begin{algorithm}[t]
 	\caption{QoS-Aware Communications with STAR-RISs}
 	\label{alg:1}
 	\begin{algorithmic}[1]
 		\REQUIRE Network topology, channel parameters
 		\STATE \textbf{Initialize:} \(n\leftarrow n+1\); feasible \(\boldsymbol{w}^{(n)}\), \(\boldsymbol{\Theta}^{{\mathsf t},(n)}\), \(\boldsymbol{\Theta}^{{\mathsf r},(n)}\); set tolerance \(\epsilon\)
            \STATE \(n\leftarrow n+1\)
 		\REPEAT
 		\STATE Update \(\boldsymbol{w}^{(n)}\) by solving~\eqref{eq:12};
 		\STATE Update \(\boldsymbol{\Theta}^{{\mathsf t},(n)}\), \(\boldsymbol{\Theta}^{{\mathsf r},(n)}\) by solving~\eqref{eq:19};
            \STATE Update \(\boldsymbol{\alpha}^{(n)}\) and \(\boldsymbol{\eta}^{(n)}\) via closed-form expressions ~\eqref{eq:7};

 		\UNTIL{Objective value satisfies \(\left|\frac{f^{(n)}-f^{(n-1)}}{f^{(n-1)}}\right|\le\epsilon\)}
 		\ENSURE Optimized \(\boldsymbol{w}\), \(\boldsymbol{\Theta}^{\mathsf t}\), \(\boldsymbol{\Theta}^{\mathsf r}\)
 	\end{algorithmic}
 \end{algorithm}

\subsection{RIS Passive Beamforming }
With the transmit beamforming vectors and auxiliary variables fixed, we optimize the passive beamforming at the STAR-RIS. For ease of analysis, we reformulate the transmission model for each user type separately. For transmission users \(U_{k, j_k}\), the effective channel can be rewritten as
\begin{equation}
		\begin{aligned}
			\boldsymbol{h}_{k,j_k}^{\mathsf H} \boldsymbol{\Theta}_k^{\mathsf t} \boldsymbol{H}_{0,k} = \left(\boldsymbol v_k^{\mathsf t} \right)^{\top} \operatorname{diag}\left(\boldsymbol{h}_{k,j_k}^{\mathsf H} \right)\boldsymbol{H}_{0,k} = \boldsymbol \vartheta_k^{\mathsf H}  \boldsymbol H_{k,j_k}
		\end{aligned}
\end{equation}
where \(\boldsymbol H_{k,j_k} = \operatorname{diag}\left(\boldsymbol{h}_{k,j_k}^{\mathsf H} \right)\boldsymbol{H}_{0,k}\) and \(\boldsymbol \vartheta_k^{\mathsf H} = \left(\boldsymbol v_k^{\mathsf t} \right)^{\top} \) are newly introduced variables. For reflection users \(u_{j_0}\), the combined reflected and direct channel is expressed as
\begin{equation}
    \begin{aligned}
& \boldsymbol{h}_{0,j_0}^{\mathsf H} + \sum_{k\in\mathcal{K}} \boldsymbol{h}_{k,j_0}^{\mathsf H} \boldsymbol{\Theta}_k^{\mathsf r} \boldsymbol{H}_{0,k} \\=& 
        \begin{bmatrix}
    \left(\boldsymbol v^{\mathsf r} \right)^{\top}&1
        \end{bmatrix}
        \begin{bmatrix}
    \operatorname{diag}\left(\boldsymbol h_{j_0}^{\mathsf H} \right)\boldsymbol H_0\\
    \boldsymbol{h}_{0,j_0}^{\mathsf H}
        \end{bmatrix} 
        = \boldsymbol \varphi^{\mathsf H}  \boldsymbol H_{j_0}
        ,
    \end{aligned}
\end{equation}
where \(\boldsymbol{h}_{j_0}^{\mathsf H} = [\boldsymbol{h}_{1,j_0}^{\mathsf H},\dots,\boldsymbol{h}_{K,j_0}^{\mathsf H}]\), \(\boldsymbol{H}_0 =  [\boldsymbol{H}_{0,1}^\top,\dots, \boldsymbol{H}_{0,K}^\top]^\top\) and \((\boldsymbol{v}^\mathsf r)^\top = [(\boldsymbol{v}_1^{\mathsf r})^\top, \dots, (\boldsymbol{v}_K^{\mathsf r})^\top]^\top\) are reorganized forms of existing variables. The newly introduced variables are \(\boldsymbol{H}_{j_0} = [\operatorname{diag}(\boldsymbol{h}_{j_0}^{\mathsf H})\boldsymbol{H}_0; \boldsymbol{h}_{0,j_0}^{\mathsf H}]\) and \(\boldsymbol \varphi^{\mathsf H} = [\left(\boldsymbol v^{\mathsf r} \right)^{\top}\;1]\). We continue applying the SDP technique by introducing the set of matrix variables \(\mathcal{V} = \{\boldsymbol{\Psi}_k\}_{k \in \mathcal{K}} \cup \{\boldsymbol{\Phi}\}\), where \(\boldsymbol{\Psi}_k = \boldsymbol{\vartheta}_k \boldsymbol{\vartheta}_k^{\mathsf H}\) and \(\boldsymbol{\Phi} = \boldsymbol{\varphi} \boldsymbol{\varphi}^{\mathsf H}\). Each matrix in \(\mathcal{\boldsymbol V}\) is subject to the constraints \(\boldsymbol{V} \succeq \boldsymbol{0},\ \text{rank}(\boldsymbol{V}) = 1,\ \forall \boldsymbol{V} \in \mathcal{V}\). Thus, the desired signal power and the total received power are given by 
\begin{equation}
P_u({\mathcal V}) = 
\begin{cases}
	\operatorname{Tr}(\boldsymbol\Psi_k\boldsymbol{\Xi}_{k,j_k}^k), & \text{if }  u = j_k, \\
	\operatorname{Tr}(\boldsymbol{\Phi}\boldsymbol{\Xi}_{j_0}^{j_0}), & \text{if }u = j_0,
\end{cases}
\end{equation}
and
\begin{equation}
Q_u({\mathcal V}) =
\begin{cases}
	\sum\limits_{i\in\mathcal{I}}\operatorname{Tr}(\boldsymbol\Psi_k\boldsymbol{\Xi}_{k,j_k}^i) + \sigma_0^2, & \text{if }u = j_k, \\
	\sum\limits_{i\in\mathcal{I}}\operatorname{Tr}(\boldsymbol{\Phi}\boldsymbol{\Xi}_{j_0}^i) + \sigma_0^2, &\text{if } u = j_0,
\end{cases}
\end{equation}
with \(\boldsymbol{\Xi}_{k,j_k}^i = \boldsymbol{H}_{k,j_k}\boldsymbol{w}_i\boldsymbol{w}_i^{\mathsf H}\boldsymbol{H}_{k,j_k}^{\mathsf H}\) and \(\boldsymbol{\Xi}_{j_0}^i = \boldsymbol{H}_{j_0}\boldsymbol{w}_i\boldsymbol{w}_i^{\mathsf H}\boldsymbol{H}_{j_0}^{\mathsf H}\). Accordingly, the objective function can be reformulated as
\begin{equation}
	\overline{f}(\mathcal{ V}) = \sum_{u\in\mathcal{U}}\left(2\eta_u\sqrt{(1+\alpha_u)P_u({\mathcal V})} - \eta_u^2 Q_u({\mathcal V})\right), \label{eq:14}
\end{equation}
\begin{figure}[t]
		\centering{\includegraphics[trim={0.1cm 0cm 0.45cm 0.3cm},clip,scale=0.8]{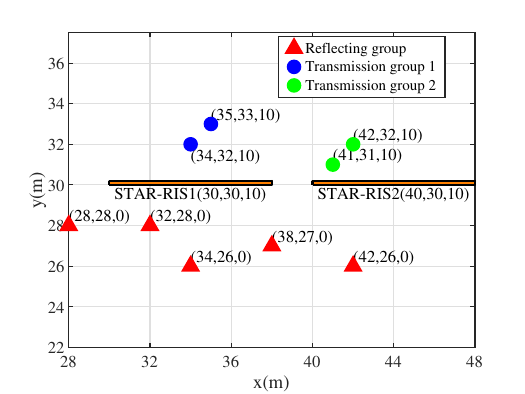}}
		\caption{Network topology for simulation.}
		\label{fig:location}
\end{figure}
Similarly, the QoS constraints are equivalently expressed in the SDP form as follows
\begin{equation}
	(2^{R_u^{\min}} - 1) Q_u(\boldsymbol{\mathcal V}) - 2^{R_u^{\min}} P_u(\boldsymbol{\mathcal V}) \leq 0\quad \forall u\in\mathcal{U}. \label{eq:15}
\end{equation}

Therefore, the RIS beamforming subproblem is formulated as
\begin{subequations}\label{eq:19}
	\begin{align}
		\max_{ {\mathcal V}} \quad & \overline{f}({\mathcal V})  \\
		{\rm s.t.} \quad
		& \boldsymbol{V} \succeq 0, \operatorname{rank}\left(\boldsymbol V \right)=1, \quad \forall \boldsymbol V \in \mathcal{ V}, \\
		& \boldsymbol{\Psi}_k^{n_k,n_k} + \boldsymbol{\Phi}^{p,p} = 1, \quad \forall n_k\in\mathcal N_k,\forall k\in\mathcal{K} \\
		& \boldsymbol{\Phi}^{\text {end},\text {end}} = 1, 
	\end{align}
\end{subequations}
where the QoS constraints \eqref{eq:15} are also included, and \(p = n_k + \sum_{m=1}^{k-1} N_m\) denotes the mapping from local STAR-RIS element indices to the global matrix index. For the phase shift design, we similarly apply semidefinite relaxation and ignore the rank-one constraint. The relaxed problem can be efficiently solved using standard convex optimization solvers such as CVX. If the obtained solution does not satisfy the rank-one condition, a feasible approximation can be generated via Gaussian randomization. The effective STAR-RIS transmission and reflection coefficients are then recovered as $v_{k,n_k}^{\mathsf{t}} = \vartheta_{k,n_k}^*$, $v_{k,n_k}^{\mathsf{r}} = \varphi_p^*$, $\forall n_k \in \mathcal{N}_k, \forall k \in \mathcal{K}$, where \( (^*)\) denotes the complex conjugate.

\subsection{Auxiliary Variables}
	
Finally, after fixing the transmit and passive beamforming, the subproblems for the auxiliary variables introduced by the LDT-QT reformulation are convex. Consequently, their closed-form updates at the iteration are derived from the first-order optimality conditions as
\begin{subequations}\label{eq:7}
	\begin{align}
		\alpha_u &= \frac{P_u}{Q_u-P_u}, \quad \forall u \in \mathcal{U}, \label{eq:7a} \\
		\eta_u &= \frac{\sqrt{(1+\alpha_u)P_u}}{Q_u}, \quad \forall u \in \mathcal{U}. \label{eq:7b}
	\end{align}
\end{subequations}

In summary, the original problem \eqref{eq:main2} is thus decomposed and solved under a BCD framework via subproblems \eqref{eq:12}, \eqref{eq:19}, and \eqref{eq:7}, as outlined in Alg.~\ref{alg:1}. This algorithm adopts a block coordinate descent framework. Since the objective function is non-decreasing at each iteration and the transmission power is bounded, the objective value has an upper bound, so the algorithm is guaranteed to converge. In each outer BCD iteration the algorithm solves three sub-problems: BS beamforming, STAR-RIS beamforming and auxiliary-variable updating. With an interior-point solver, the BS-beamforming sub-problem has complexity $\mathcal{O}((K + J_0)^{1.5}N_0^{3.5})$, while the STAR-RIS beamforming sub-problem requires $\mathcal{O}((\sum_{k=1}^{K}N_k + 1)^{4.5})$. The auxiliary-variable step is obtained in closed form and its cost is negligible. Consequently, if the algorithm converges within $L$ iterations, the total computational complexity is $\mathcal{O}\left(L\left((K + J_0)^{1.5}N_0^{3.5} + (\sum_{k=1}^{K}N_k + 1)^{4.5}\right)\right)$.

	\section{Simulation Results}

		\begin{figure*}[t]
		\centering
		
		\begin{subfigure}{0.32\textwidth}
			\centering
			\includegraphics[trim={0.1cm 0cm 0.45cm 0.3cm},clip,width=\textwidth]{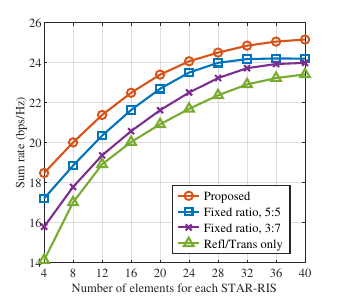}
			\caption{ }
			 \label{fig:subfig3a}
		\end{subfigure}
		\hfill
		\begin{subfigure}{0.32\textwidth}
			\centering
			\includegraphics[trim={0.1cm 0cm 0.45cm 0.3cm},width=\textwidth]{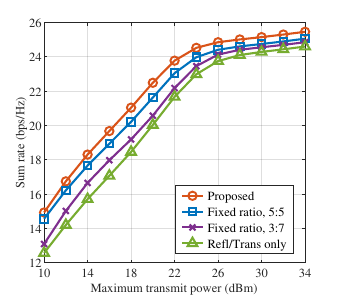}
			\caption{ }
			 \label{fig:subfig3b}
		\end{subfigure}
		\hfill
		\begin{subfigure}{0.32\textwidth}
			\centering
			\includegraphics[trim={0.1cm 0cm 0.45cm 0.3cm},width=\textwidth]{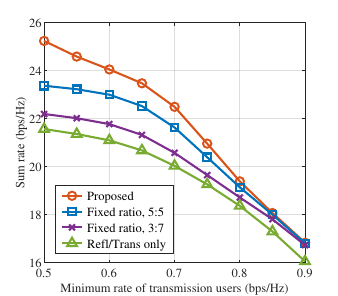}
			\caption{ }
			 \label{fig:subfig3c}
		\end{subfigure}
		
		\caption{System performance comparison. (a) Sum rate vs number of elements in the STAR-RIS. (b) Sum rate vs maximum transmit power. (c) Sum rate vs minimum rate of transmission users. }
		\label{fig:figure3}
	\end{figure*}

    This section presents simulation results for the proposed multi-STAR-RIS-assisted system. As illustrated in \Cref{fig:location}, the setup includes 2 STAR-RISs and multiple users deployed in space. A BS equipped with 5 antennas is located at \((0, 0, 20)\) meters, and each STAR-RIS comprises 16 passive elements.Three user groups are considered: (i) a reflection group with 5 outdoor users, and (ii) two transmission groups, each served by one STAR-RIS and containing 2 indoor users. The BS transmits with a maximum power of \(0.1~\text{W}\).  To guarantee diversified QoS, a 2:1 rate ratio is maintained between the reflection and transmission users, where $R_1$ and $R_2$ denote their respective rate requirements. Specifically, the minimum rate requirements are set to $1.4~\text{bps/Hz}$ and $0.7~\text{bps/Hz}$, respectively. Large-scale fading follows a free-space path loss model with \(-20~\text{dB}\) attenuation at 1 meter, and the noise power is uniformly set to \(-140~\text{dBW}\). The convergence threshold for the iterative algorithm is \(10^{-3}\). Three baseline schemes are considered for comparison: 1) \textbf{Fixed ratio (5:5)}, each STAR-RIS element equally splits energy between reflection and transmission (\(\beta_{k,n_k}^{\mathsf{r}} = \beta_{k,n_k}^{\mathsf{t}} = 0.5\)); 2) \textbf{Fixed ratio (3:7)}, 30\% energy allocated to transmission and 70\% to reflection (\(\beta_{k,n_k}^{\mathsf{t}} = 0.3\), \(\beta_{k,n_k}^{\mathsf{r}} = 0.7\)); 3) \textbf{Refl/Trans only}, elements are alternately configured: odd-indexed ones in reflection mode (\(\beta_{k,n}^{\mathsf{r}} = 1\)), even-indexed ones in transmission mode (\(\beta_{k,n}^{\mathsf{t}} = 1\)).

	\Cref{fig:subfig3a} illustrates the total system rate versus the number of elements per STAR-RIS. All schemes benefit from increased RIS elements, confirming the spatial degree-of-freedom gain brought by STAR-RISs. The proposed method consistently achieves the highest total rate across the entire range, outperforming benchmark schemes including fixed ratio (5:5, 3:7) and Refl/Trans only designs. This demonstrates its superior capability in leveraging additional STAR-RIS elements. When $N \geq 30$, the performance gain saturates, indicating diminishing marginal returns.
	
        \Cref{fig:subfig3b} shows how the system sum rate varies with the maximum transmit power. As $P_T$ increases, the performance of all schemes improves significantly, indicating that the enhanced signal power yields effective link gains. In the region where $P_T < 24\,\text{dBm}$, the sum rate experiences rapid growth, being primarily limited by background noise. Conversely, in the high-power region, this growth decelerates and the sum rate tends to saturate, mainly due to the impact of multi-user interference. These results further validate the critical role of the joint optimization of transmit beamforming  and passive beamforming  in enhancing the system  performance.

        \Cref{fig:subfig3c} shows how the system sum rate varies with the minimum rate threshold for transmission users. It can be observed that with the increase of threshold, the system sum rate of all schemes gradually decreases, reflecting that more stringent QoS guarantees reduce the overall schedulability of resources. When threshold approaches its maximum value, the performance of the proposed scheme becomes close to that of the fixed-ratio schemes. This suggests that under highly demanding QoS requirements, the optimization margin is limited, and simpler configurations can achieve comparable performance. Notably, however, both the proposed and the fixed-ratio schemes (which are STAR-RIS based) consistently and significantly outperform the ``Refl/Trans only" design across the entire $R_2$ range. This further validates the structural advantages of the STAR-RIS architecture in complex QoS provisioning scenarios.
    
	\section{Conclusion}

    This study proposed a QoS-aware communication framework for multi-STAR-RIS-assisted multi-user systems. To accommodate heterogeneous user requirements, we jointly optimized the BS transmit beamforming and the STAR-RIS transmission/reflection coefficients under group-specific rate constraints. An efficient BCD-based algorithm was developed, leveraging Lagrangian duality and quadratic transformation to handle the non-convex optimization. Simulation results demonstrated that the proposed approach significantly improves the aggregate link rate while ensuring diversified QoS guarantees. These results offer practical insights for the design of future STAR-RIS-enabled wireless networks.

\printbibliography 
	
\end{document}